\documentclass[journal]{IEEEtran}
\usepackage{caption}
\usepackage{graphicx}
\usepackage{bm}
\usepackage{amsmath}

\ifCLASSINFOpdf
\else
\fi

\usepackage{xcolor}

\begin{document}

\title{Enhancement of Satellite-to-Phone Link Budget by Using Distributed Beamforming}

\author{Zhuoao~Xu,
        Yue~Gao,~\IEEEmembership{Senior~Member,~IEEE,}
        Gaojie~Chen,~\IEEEmembership{Senior~Member,~IEEE,}
        Ryan~Fernandez,
        Vedaprabhu~Basavarajappa,~\IEEEmembership{Senior~Member,~IEEE,}
        and~Rahim~Tafazolli,~\IEEEmembership{Senior~Member,~IEEE}
\thanks{Z. Xu, Y. Gao, G. Chen, R. Fernandez, V. Basavarajappa and R. Tafazolli are with the Institute for Communication Systems (ICS), 5GIC \& 6GIC, University of Surrey, Guildford, Surrey, GU2 7XH, U.K. Email:\{z.xu, yue.gao, gaojie.chen, r.fernandez, v.basavarajappa, r.tafazolli\}@surrey.ac.uk.}
}

\maketitle

\begin{abstract}
Small satellites in Low Earth Orbit (LEO) attract much attention from both industry and academia. The latest production and launch technologies constantly drive the development of LEO constellations. However, the wideband signal, except text messages, cannot be transmitted directly from an LEO satellite to a standard mobile cellular phone due to the insufficient link budget. The current LEO constellation network has to use an extra ground device to receive the signal from the satellite first and then forward the signal to the User Equipment (UE). To achieve direct network communications between LEO satellites and UE, we propose a novel distributed beamforming technology based on the superposition of electromagnetic (EM) waves radiated from multiple satellites that can significantly enhance the link budget in this paper. EM full-wave simulation and Monte Carlo simulation results are provided to verify the effectiveness of the proposed method. The simulation results show a nearly 6 dB enhancement using two radiation sources and an almost 12 dB enhancement using four sources. The received power enhancement could be doubled compared to the diversity gain in Multiple-Input and Single-Output (MISO). Furthermore, other practical application challenges, such as the synchronization and Doppler effect, are also presented.

\end{abstract}


\IEEEpeerreviewmaketitle

\section{Introduction}

\IEEEPARstart{I}{n} the 1990s, Low Earth Orbit (LEO, less than 2000 km above the Earth's surface) constellations such as `Iridium' and `Globalstar' emerged and aimed to provide portable satellite phone service with global coverage. Still, most of them could not sustain their development due to their big rival, the 3G terrestrial cellular network, which can provide better performance with lower costs.

However, over the past few years, LEO constellations have staged a comeback because of the rapid development of satellite technologies, increased demand, and much lower launch costs. Compared to the old LEO constellations, new LEO constellations have a much better performance which benefits from their digital communication payloads, advanced modulation schemes, multi-beam antennas, and more sophisticated frequency reuse schemes \cite{LEO3}. Among them, the most crucial factor is launch technology. Rocket reuse and launching multiple satellites with one rocket have significantly reduced launch costs \cite{6GVision}. New LEO constellation representatives, Starlink and OneWeb, launched their first test satellites in 2018 and 2019, respectively. 

Standard mobile cellular phones do not have enough reception gain, and the LEO satellite's transmit power, and antenna gain is limited by size and weight, resulting in an inadequate link budget for direct wideband communication between LEO satellites and User Equipment (UE). Beam link budgets for the user downlink for three LEO constellations were given in \cite{LEO3}. Take the Starlink constellation as an example, although it reduces the orbital altitude from 1150 km to 550 km, the received power is still less than -110 dBm following the link budget calculation process in \cite{LinkBudget}. Given this, the current solution adopted by Starlink is using an extra device comprised of a phased antenna array with a high gain of more than 30 dBi. The device receives signals from the LEO satellite and then forwards the signals to UE, which limits the portability of the LEO satellite networks. A few advanced companies, such as LYNK, AST, and SpaceX, are exploring a direct way to connect cellular phones with LEO satellites. However, with limited bandwidth, initially, their service will offer text messages only \cite{LYNK}. Therefore, how to realize a future LEO satellite network for direct access from a mobile cellular phone has become a hot issue.

Recent work by Mohammed \cite{CF-mMIMO} proposed a Cell-Free massive Multiple-Input Multiple-Output (CF-mMIMO) based architecture and discussed various aspects of ultra-dense LEO satellite networks design, but mainly focused on a joint optimization framework for the power allocation and handover management processes. The MIMO applicability to Ultra-High Frequency (UHF) satellite communications in the geostationary orbital space was analyzed in \cite{SatMIMO 16}. The results show that applying the MIMO technique can increase the channel capacity in narrowband UHF satellite communications. The authors in \cite{SatMIMO 21} considered a Land Mobile Satellite (LMS) MIMO, where two Geosynchronous Earth Orbit (GEO) satellites simultaneously communicate with a mobile User Terminal (UT). Its simulation results show that dual-satellite MIMO communications can achieve better Bit Error Rate (BER) performance under the same Signal-to-Noise-Ratio (SNR) condition compared to single-satellite communications. However, the diversity or multiplexing gain provided by Multiple-Input and Single-Output (MISO) is only proportional to the number of coordinated satellites $N$ \cite{MISO}. To further improve the received signal strength, we propose a new distributed beamforming technology. Compared to the maximum diversity gain of $N$ in MISO, the received power could be enhanced by $N^2$ times at maximum. As known to all, beamforming is an effective technology in modern communication systems that can concentrate energy and extend the communication range \cite{Beamforming}. It then evolved to distributed transmit beamforming \cite{DTB}. However, the distributed beamforming applied to LEO constellations is significantly distinct from that for terrestrial networks. The main reason is that the satellites are so far away from each other that they cannot form a main beam as in conventional arrays. Another reason is that satellites keep moving along their orbits, causing their distribution changes at any time.

The rest of this article is organized as follows. Section II gives a link budget calculation based on 5G UE operating frequency. The distributed beamforming technology is then proposed to compensate for the insufficient link budget. Both the structure of distributed array and the theory are presented in detail. Then, in Section III, Simulations are given to verify that received power could be enhanced through distributed beamforming, and the beam coverage patterns could be designed by changing relevant parameters. Afterward, we introduce a few challenges needed to be addressed in the future in Section IV. Finally, the conclusion and future works will be given in Section V.

\section{A Novel Distributed Beamforming for LEO Satellite-based Networks}

\subsection{Link Budget}
We provide the link budget in Table I to achieve direct 5G communication between the LEO satellite and UE. Based on the current 5G network and smartphones working at 3.5 GHz, the received power of the signal transmitted from a single satellite is calculated as -100.4 dBm. However, from 3GPP TS 38.101-1 \cite{3GPP}, it is known that the minimum reference sensitivity for operating band n78 is -96.5 dBm. Thus, to achieve direct LEO satellite network access from UE, the solution must at least provide a 4 dB enhancement. As the demand for bandwidth increases, more received power will be required. For example, 100 MHz bandwidth corresponds to 13.5 dB power improvement.

\begin{table}[t]
 \centering
 \caption{Link Budget.} 
     \begin{tabular}{c|c}

    \hline
     \hline
       \textbf{ Parameters}  &  \textbf{ Values}  \\
        \hline
         \hline
         
       Distance  &  550 km \\
    \hline
       Operating frequency  &  3.5 GHz \\
    \hline
       Free space path loss (FSPL)  &  158.1 dB \\
    \hline
       Effective isotropic radiated power (EIRP)  &  36.7 dBW \\
    \hline
       Transmit antenna gain  &  37.1 dBi \\
    \hline
       Receive antenna gain  &  0 dBi \\
    \hline
       Atmospheric losses and Rain attenuation  &  5 dB \\
    \hline
       Losses from transmitter  &  2 dB \\
    \hline
       Losses from receiver  &  2 dB \\
    \hline
       Received power calculated & -100.4 dBm \\
    \hline
     \end{tabular}
 \end{table}
 

\subsection{System Model of Distributed Beamforming}

\begin{figure}
  \centering
  \includegraphics[width=3.3in]{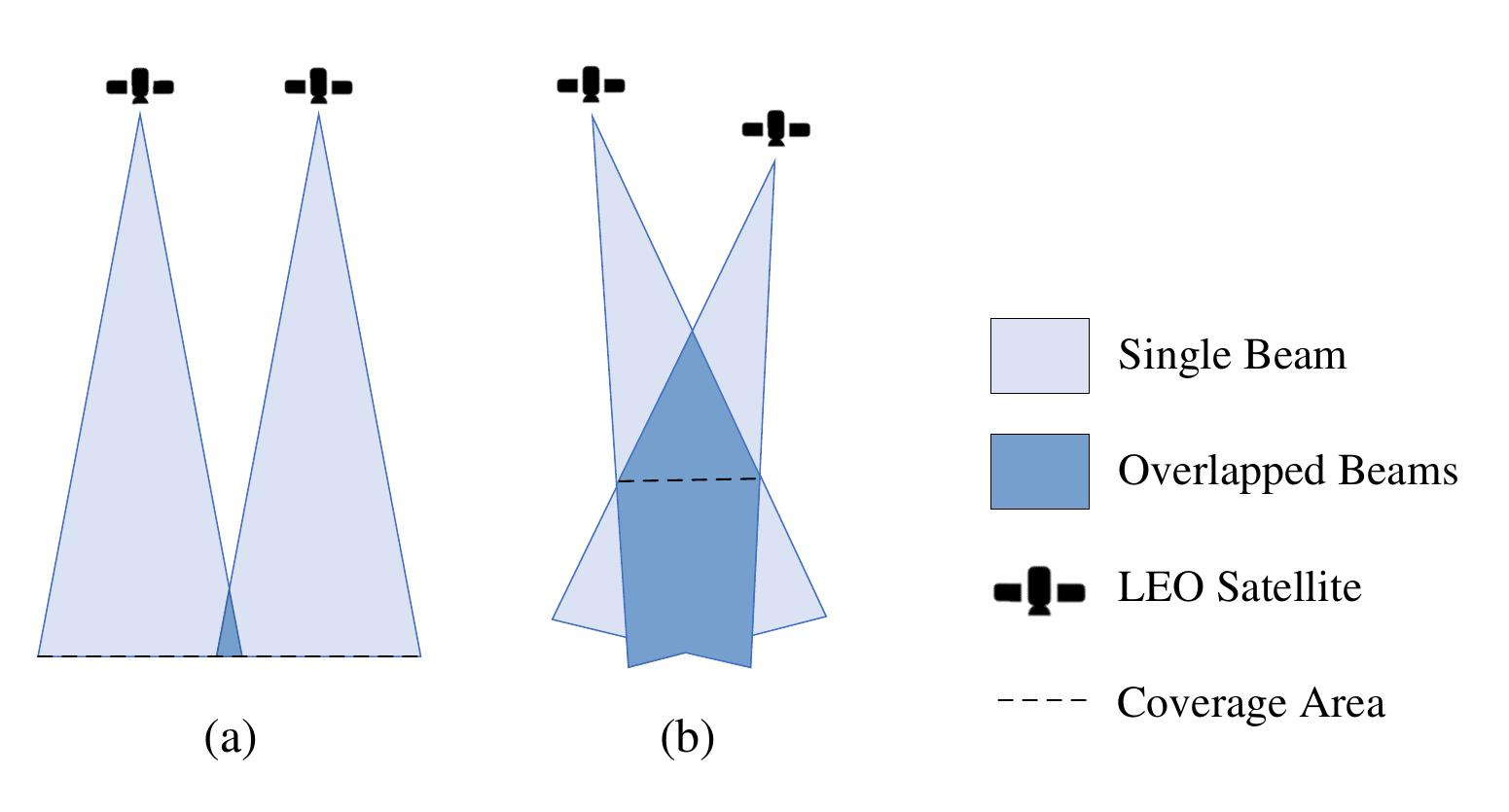}
  \caption{Beam coverage scheme. (a) Satellites work independently and cover evenly. (b) Same coverage by distributed beamforming.}\label{fig1}
\end{figure}

As we know, the LEO mega constellation comprises tens of thousands of small satellites that run in fixed orbits with different altitudes, and phased array antenna is commonly used on satellite which can provide high gain and electrically steerable beam scanning. The current coverage scheme is shown in Fig. 1 (a), where each satellite has its coverage area and works independently. When the UE is nearly moving out, the coverage of the present satellite is limited by the elevation angle; the satellite should smoothly hand over the UE to the incoming satellite. To achieve seamless coverage, inevitably, there will be some overlap at the edge of the beam, but the interference within the overlapping area is unwanted. Inversely, for our distributed beamforming technology, constructive interference is exactly what is desired. The authors in \cite{Superposition} present research on how Electromagnetic (EM) waves interact and exchange energy. From the article, it can be found that beams merely interfere constructively or destructively in the overlapped area and remain in their original propagation after passing it. Under this circumstance, satellites need to steer their radiation beams properly for accurate coverage. As shown in Fig. 1 (b), the dashed line represents beams radiated from satellites propagating towards the same coverage area.

\begin{figure}
  \centering
  \includegraphics[width=2.5 in]{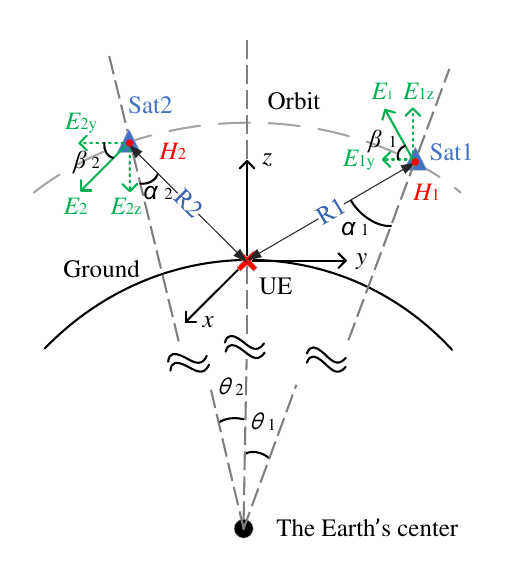}
  \caption{The superposition of EM waves radiated from LEO satellites. The black solid curve represents the ground, the grey dashed curve above it represents the satellite operating orbit, and the black point below represents the Earth's center.}\label{fig2}
\end{figure}

Fig. 2 depicts the EM waves superposition principle. Assuming Satellite 1 (Sat1) and Satellite 2 (Sat2) run in the same orbit, communicate with the same UE on the ground and are located on either side of the center line. In a spherical coordinate system with the Earth's center as the origin, the central angles of Sat1 and Sat2 are $\theta_1$ and $\theta_2$, respectively. $\alpha_1$ denotes the angle between Sat1's line with the Earth's center and its line with the UE, and so does $\alpha_2$. As known to all, radiating EM waves are Transverse Electromagnetic (TEM) waves. Thus, The electric (E) field vector, the magnetic (H) field vector, and the wave vector (W) are perpendicular to each other. Considering there is also a rectangular coordinate with the UE as the origin, E fields and H fields can be decomposed along the y \& z-axis and the x \& z-axis independently. $\beta_1$ and $\beta_2$ represent the decomposition angles of $E_1$ and $E_2$. According to the geometry, it can be easily obtained that $\beta_1 = \alpha_1 + \theta_1$, similarly, $\beta_2 = \alpha_2 + \theta_2$. Besides, $R1$ and $R2$ are the distance from Sat1 and Sat2 to the UE.

\subsection{Performance Analysis}

As shown in Fig. 2, supposing the two satellites above the UE work coherently. In other words, two beams of the same frequency arrive at the receiver simultaneously and in phase, then the E field components along the $x$-axis would be constructively superimposed. At the same time, the E field components along the $z$-axis should be subtracted herein while they should be summed when they are in the same space divided by the $xoz$ plane. The superposition of the H field is similar, except for the substitution of $x$ for $y$. But in Fig. 2, due to the two H fields having the same direction, they can be directly added together. After superimposing the E and H fields in their decomposition surfaces, synthesized E and H vectors could be obtained. The average energy flux (power per unit area) can be further calculated by average Poynting vector calculation as follows,
\begin{equation} \label{eq1}\small
\begin{split}
\bm{S_{av}}&=\frac{1}{T}\int^{T}_{0}\sum_{i=1}^{2}(\bm{E_{iy}}+\bm{E_{iz}})\times\sum_{i=1}^{2}\bm{H_{i}}\rm{d} \textit{t}\\
&=\frac{E_{0}^{2}}{Z_{0}}[(\cos\beta_{1}+\cos\beta_{1}\cos\Delta\varphi+\cos\beta_{2}\cos\Delta\varphi+\cos\beta_{2})\\
&\quad\ (-\bm{e_{z}})+(\sin\beta_{1} + \sin\beta_{1}\cos\Delta\varphi-\sin\beta_{2}\cos\Delta\varphi-\sin\beta_{2})\\
&\quad\ (-\bm{e_{y}})]\\
& \le \frac{2E_{0}^{2}}{Z_{0}}[(\cos\beta_{1}+\cos\beta_{2})(-\bm{e_{z}})+(\sin\beta_{1}-\sin\beta_{2})(-\bm{e_{y}})]\\
& \le \frac{4E_{0}^{2}}{Z_{0}}(-\bm{e_{z}}),
\end{split}
\end{equation}
where $T$ represents a time period, both $E_1$ and $E_2$ have the same E field effective value $\sqrt{2}E_0$, $Z_0$ is the impedance of free space, and $\Delta\varphi$ is the phase difference of incident beams caused by the routes. It can be seen from (1) that in order to approach the maximum received power, both $\Delta\varphi$ and $\beta$ should be as close to 0 as possible. Different orbital altitudes make it possible for satellites to get closer when a satellite runs right above another. Ideally, two satellites would yield four times as much received power as one satellite, meaning a 6 dB enhancement. By contrast, the received power obtained by MISO is the combination of received powers from multiple sources. For example, two satellites can provide a maximum gain of 3 dB with MISO. 

Assuming there are $N$ satellites running in the same direction and working constructively, and other conditions are the same as above, then (1) should be modified as

\begin{equation} \label{eq2}\small
\begin{split}
\bm{S_{av}}&=\frac{1}{T}\int^{T}_{0}\sum_{i=1}^{N}(\bm{E_{iy}}+\bm{E_{iz}})\times\sum_{i=1}^{N}\bm{H_{i}}\rm{d} \textit{t}\\
& \le \frac{NE_{0}^{2}}{Z_{0}}(\cos\beta_{1}+\cos\beta_{2}+...+\cos\beta_{N})(-\bm{e_{z}})\\
& \le \frac{N^{2}E_{0}^{2}}{Z_{0}}(-\bm{e_{z}}).
\end{split}
\end{equation}
The received power would ideally be increased by $N^2$ with distributed beamforming according to (2), while only a factor of $N$ can be achieved by MISO. If the increased received power is expressed in dB, the former would be twice as large as the latter. As the number of cooperating satellites increases, the received power will be significantly boosted. Finally, the received signal would become strong enough to achieve direct network communication between LEO satellites and UE. Whereas, considering the satellite spacing and inclination angle, the maximum value $N^2$ may not always be available in reality, and it will decrease when the satellite spacing and inclination angle increase.

The orbits of satellites could cross with each other. In order not to lose generality and convenience, two intersecting orbits are considered. Supposing that one orbit is along with the $x$-axis and the other orbit intersects it with an angle of $\xi$. $M$ of a total $N$ satellites move in the intersecting orbit. The average Poynting vector can be calculated as follows,
\begin{equation} \small
\begin{split}
\bm{S_{av}}&=\frac{1}{T}\int^{T}_{0}\sum_{i=1}^{N}(\bm{E_{ix}}+\bm{E_{iy}}+\bm{E_{iz}}) \times\sum_{i=1}^{N}(\bm{H_{ix}}+\bm{H_{iy}})\rm{d} \textit{t}\\
& \le \frac{[N^{2}-MN(1-\cos\xi)]E_{0}^{2}}{Z_{0}}(-\bm{e_{z}}).
\end{split}
\end{equation}
From the result, it can be known that the maximum enhancement provided by $N$ satellites in two intersecting orbits is $N^{2}-MN(1-\cos\xi)$. Similar to the previous scenario, the maximum is also obtained when $\Delta\varphi$ and $\beta$ are 0. In particular, when half of $N$ satellites move in the perpendicular orbit, the maximum enhancement will be $\frac{N^{2}}{2}$.


\section{Simulation Results and Discussion}

\subsection{EM Simulations for Superposition}

\begin{figure}
  \centering
  \includegraphics[width=3.3 in]{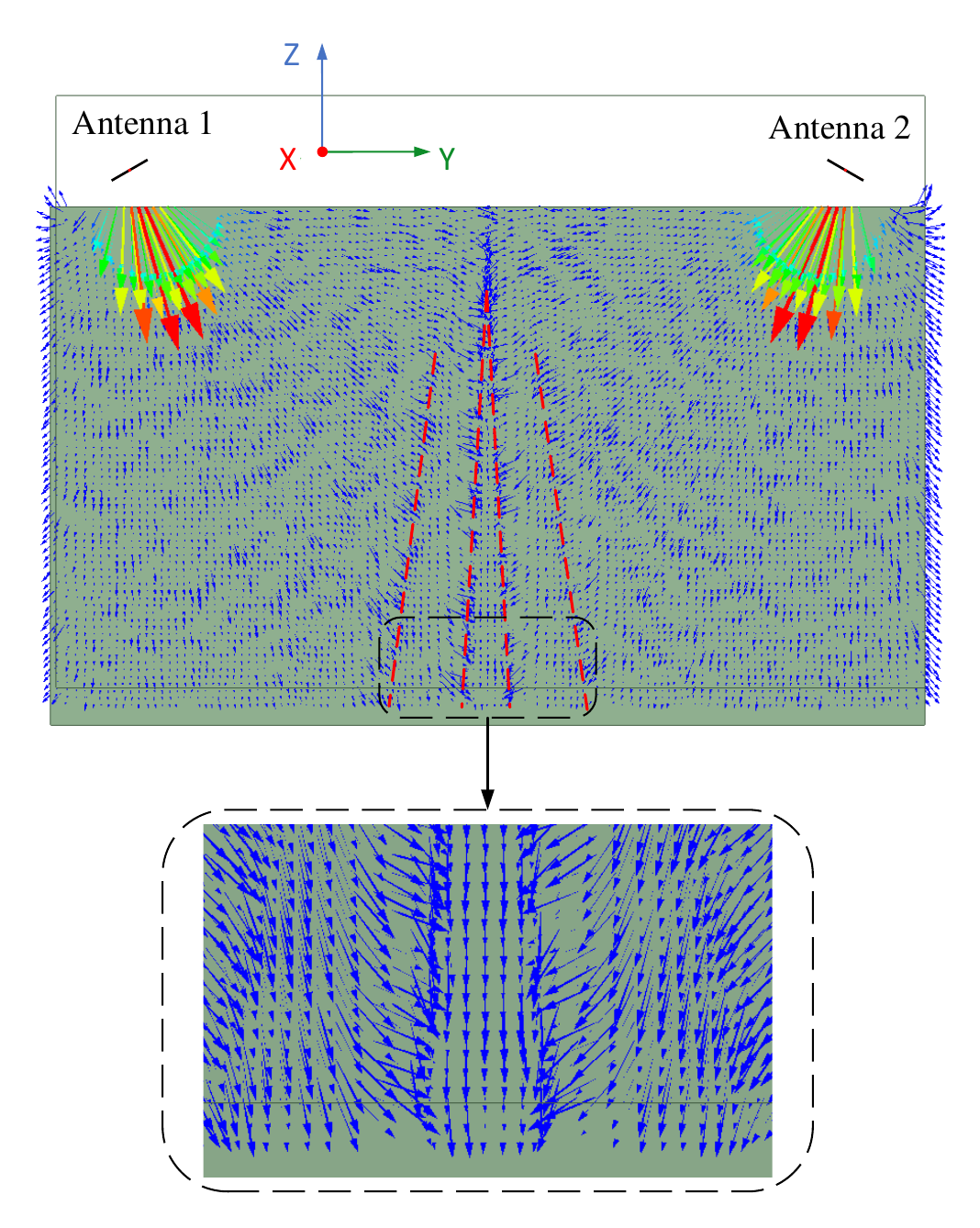}
  \caption{Two EM waves interfere constructively represented by Poynting Vector. }\label{fig3}
\end{figure}

EM simulations will be given to verify the superposition of EM waves radiated from distributed sources far apart. The radiation beam of the antenna on a satellite is usually narrow so that energy can be concentrated in the expected direction. For the convenience of verification, a dipole antenna is adopted herein without loss of generality. In particular, the simplest case with only two radiation sources is considered. 

As shown in Fig. 3, two dipole antennas are displayed symmetrically along the center line. Both have an inclination angle with a value of 30 degrees. Considering that the main beam radiation direction is perpendicular to the antenna, the main beam convergence location can be easily obtained according to the geometric structure. The operating frequency is set to 3.5 GHz, corresponding to the wavelength of 85.7 mm in a vacuum. Generally, a distance of more than ten wavelengths can be regarded as far enough for antennas to work independently. Therefore, the two half-wavelength dipoles are displayed ten wavelengths apart. 
Besides, each source's input power and initial phase can be modified based on requirements. Given the geometry of the model, constructive interference will appear along the center line when the initial phases and power are set to the same. 


In addition, Poynting Vector represented by arrows is plotted on the green rectangular sheet. The arrow size and color indicate its value, and the arrow's point indicates its direction. It can be easily found that radiation power gets weaker as it propagates away. Significantly, the synthetic Poynting vector around the center line points straight down because the transverse components cancel each other. The red dashed line represents the destructive interference, where the dominant longitudinal components cancel each other. From the point of view of vector superposition, the synthetic vector would become more prominent when the angle between vectors is less than 90 degrees. As a result, the received power can be enhanced due to the constructive interference brought by distributed beamforming.

\begin{figure}
  \centering
  \includegraphics[width=3.4in]{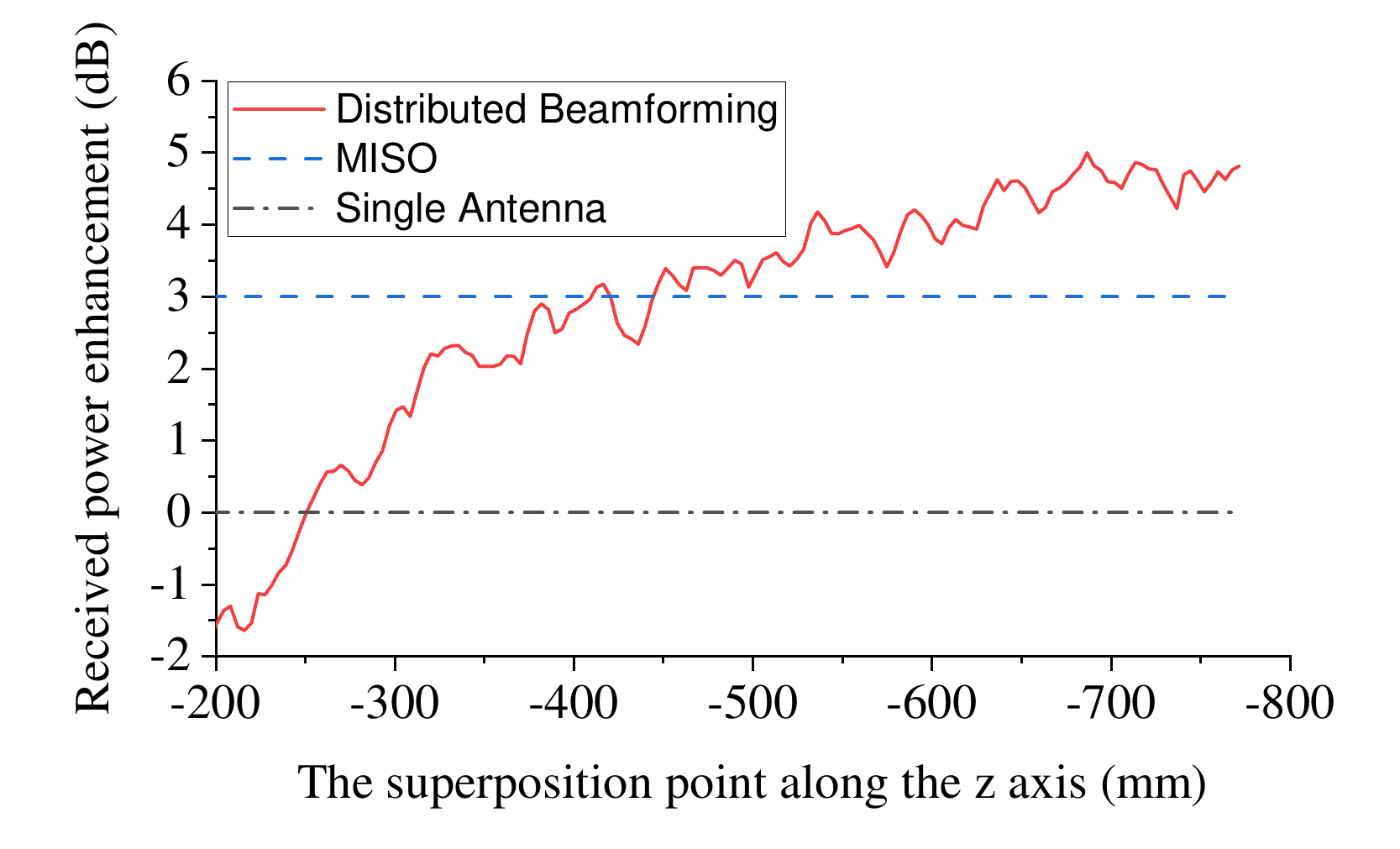}
  \caption{Comparison of the received power at different positions along the center line by three methods.}\label{fig4}
\end{figure}

As mentioned above, the same setup and symmetrical position of the two sources ensure that their radiation waves constructively interfere with each other along the center line. The synthetic Poynting vector on the line is extracted from $z$ = -200 mm to $z$ = -770 mm to compare with that of a single antenna and that of MISO. Fig. 4 illustrates the power received in three ways: one satellite, two satellites by MISO, and two satellites by distributed beamforming. As a reference, the received power of the single radiation source is normalized. The received power obtained through MISO is proportional to the number of sources as depicted in Section II, increasing the power twice or 3 dB in decibels. By contrast, the power achieved by distributed beamforming is even lower than that of the single antenna at the beginning, but it surpasses that of MISO as the observation point gets close to the target point (around $z$ = -742 mm). Under the condition of a 30-degree inclination, the received power enhancement is around 4.7 dB at the target. Through the EM simulations, the possibility of increasing the received power by distributed beamforming is verified.


\subsection{Coverage Pattern}

The receiver gains from constructive interference, accordingly, the surroundings should lose some power due to the destructive interference. Since the distance between the satellites is much longer than the wavelength, there may be frequent switching between constructive and destructive interference within the coverage. Therefore, knowing the beam coverage pattern generated by interference is vital. We provide pattern simulations obtained under different input parameters. As mentioned in the previous part, the initial variables include transmitting power, initial phase, orbit height, and the inclination angle of the beam relative to the vertical line direction. The vertical line direction represents the direction in which the satellite points towards the Earth's center.

Assuming that the single beam width of the satellite is 2.5 degrees, and given that the operating satellite height is 550 km, it can be concluded that the coverage area of the single beam is a circle at nadir with a radius of 12.5 km. The power density at boresight is the largest and drops to half the maximum value at the edge, which is -3 dB. As the beam slants away from the nadir, the coverage pattern gradually changes from a circle to an ellipse. Moreover, as the transmitting beam is steered, the power is adjusted to maintain a constant Power Flux-Density (PFD) at the Earth's surface, compensating for variations in antenna gain and path loss associated with the steering angle.

\begin{figure*}
  \centering
  \includegraphics[width=7in]{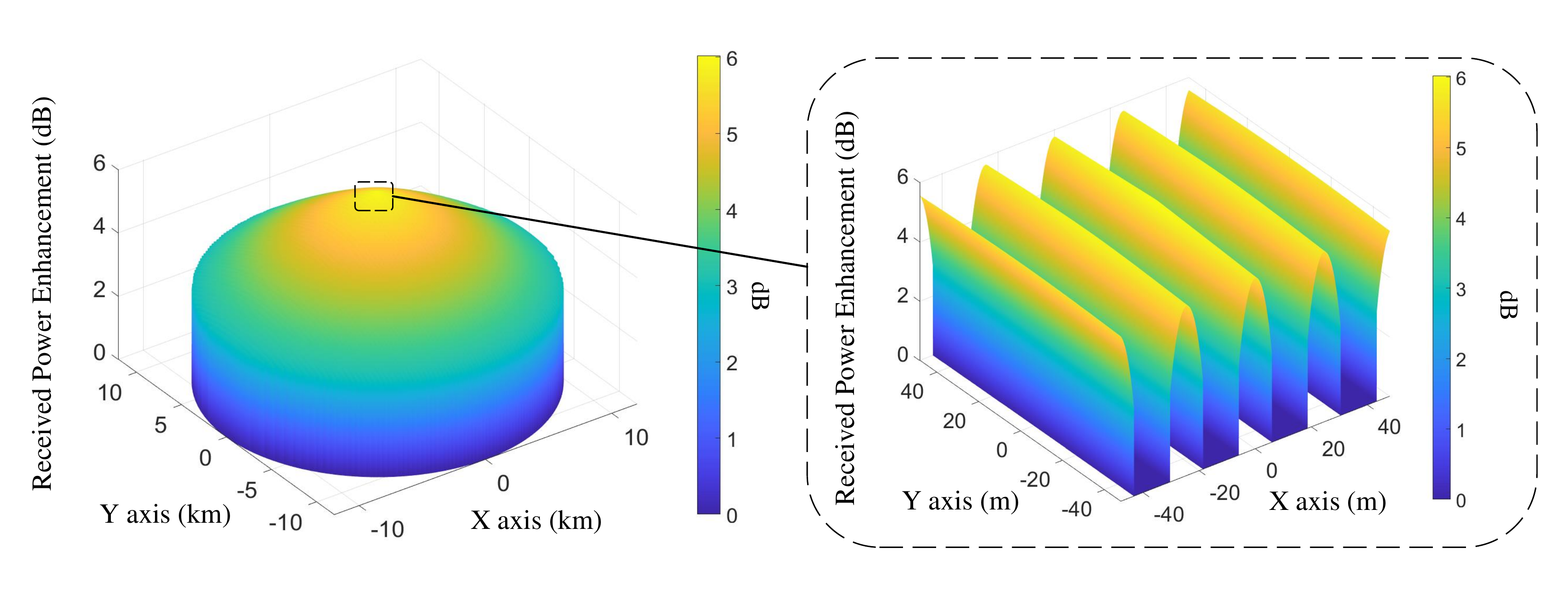}
  \caption{Coverage pattern generated by two satellites working coherently.}\label{fig5}
\end{figure*}

Here we only present two representative beam coverage cases. In the first case, only two satellites are used to implement distributed beamforming, then interference fringes appear. In cartesian coordinates, if the line between the two satellites is parallel to the X-axis, then the interference fringes will be parallel to the Y-axis. The fringe width depends on the angle of the incident wave at the receiver. When the angle between the two beams is large, the fringes are narrow and dense. On the contrary, as the angle becomes smaller, the fringes become wide and sparse. As shown in Fig. 5, when the inclination angles of both transmitting beams are 0.1 degrees, the interference fringes are broad, and the width is about 12 m. In addition, the lateral distance between the two satellites is approximately 1.92 km, which is too close for two satellites in the same orbit to achieve. But considering the high density of LEO mega constellation deployments and satellites that could cooperate with others running in different orbits, this assumption about the inclination angle is reasonable. Furthermore, it can be observed from the figure that by adopting distributed beamforming, the maximum received signal gain of nearly 6 dB can be achieved with two transmitting sources, which is larger than the 3 dB of diversity gain.

\begin{figure*}
  \centering
  \includegraphics[width=7in]{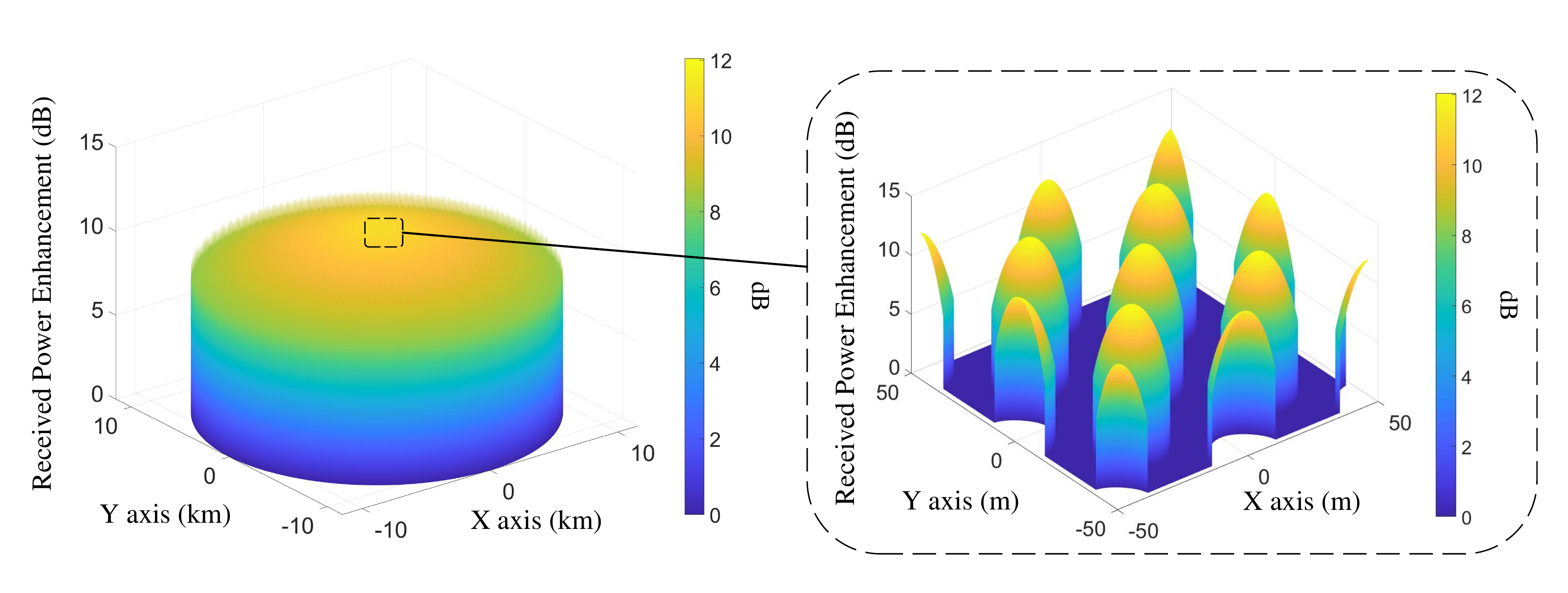}
  \caption{Coverage pattern generated by four satellites working coherently.}\label{fig6}
\end{figure*}

In the second case, four satellites are used to implement distributed beamforming, then the spot beam coverage pattern appears. More than two satellites generally create a two-dimensional distribution, resulting in spot beams. When all the inclination angles of the beams from the four directions are 0.1 degrees, interference spots will be generated whose diameters are around 24 m. 
As shown in Fig. 6, four transmitting sources could improve nearly 12 dB by distributed beamforming, which is larger than that of MISO (6 dB). The two cases above present feasible beam coverage patterns with multiple satellites working coherently. The pattern can be designed based on specific requirements by changing the initial parameters.


\section{Challenges}
To achieve portable communication between LEO satellites and UE directly, some other significant challenges would be further addressed.

\subsection{Synchronization}
In practice, the radiated EM waves must meet some synchronous characteristics to realize the constructive interference of satellite radiation beams. The first is frequency synchronization, which requires excellent stability and accuracy of satellite-mounted Radio Frequency (RF) modules. The crystal oscillators and front-end devices on different satellites need to ensure that the transmitted EM waves have the same frequency, which is the premise that multiple EM waves can be superimposed to form a standing wave. The second is phase synchronization. The incident EM waves at the receiver need to have the same phase to maximize the effect of field strength superposition. Otherwise, it may make the superimposed signal worse than the single one. Therefore, the initial phase of the transmitter needs to be carefully set after considering many factors. The third is time synchronization. Satellites are always in motion, constantly changing their positions, so the distance between each satellite and the UE is different. If they send signals to the UE simultaneously, the receiver will receive them at different times, or the signal may even be too weak to be received successfully. Thus, the transmitted signals must be set with further delays to ensure they arrive at the receiver simultaneously.

\subsection{Inter-Satellite Links}
Inter-Satellite Links (ISLs) are communication links between two or more satellites in orbit around the Earth. In an LEO constellation network, ISLs are a critical technology that enables the exchange of data and control signals between satellites in the constellation \cite{Optical}. SpaceX successfully tested its laser ISLs in late 2020 and has launched its satellites featuring laser ISLs since September 2021. Leveraging the development of ISLs and LEO constellations, the proposed approach is more feasible and does not incur much additional cost. ISLs are highly robust and can be independently networked without relying on the terrestrial network, expanding the coverage of the communication system. Additionally, their features such as high data rates and low latency can help to address the synchronization problem mentioned above. However, the acquisition, pointing, and tracking mechanisms among satellites are pretty complicated, and laser link is greatly affected by space illumination and other factors. For mega constellations, the routing of ISLs is a highly complex problem due to the relative position of LEO satellites changing all the time. In addition, signal processing on satellites dramatically increases the complexity and development difficulty, reducing the adaptability of satellites to technology upgrades and updates. This is regarded as the biggest issue that hinders the development of laser links. Overall, ISLs are a key feature of LEO constellation networks, offering a range of benefits and challenges that must be carefully considered in the design and operation of such systems.

\subsection{Doppler Effect}
Since the LEO satellites run in low-altitude orbits, they have to travel at high speed, which will change the frequency of the signal at the receiver, namely the Doppler effect. When the transmitter and receiver are getting close to each other, the frequency increases; conversely, when they move far away, the frequency decreases. The significant Doppler frequency shift increases the difficulty of receiver demodulation and degrades communication performance. \cite{Doppler} presents the analytic derivation of the Doppler shift about the signal transmitted from LEO satellites to UE. There are three main solutions for Doppler frequency shift estimation and compensation. One of them is using the geometric analysis method directly to calculate the relative velocity of communication satellites and UE and then calculate the Doppler frequency shift. Another is using the Kalman filter frequency shift estimation algorithm. The third one uses the maximum likelihood estimation algorithm to calculate the Doppler frequency shift factor, which is fed back to the frequency compensation module. Then the frequency compensation module pre-compensates the transmitted signal frequency to achieve the purpose of frequency synchronization between the transmitter and the receiver. These Doppler shift estimation and compensation methods can reduce the Doppler shift value and improve communication quality. But the reality is much more complex than the model assumes, and the specific situation should be analyzed according to the actual constellation.

\subsection{Constellation Design and Beam Switching}
Constellation design is often ambiguous for researchers because the design is constantly being adjusted as requirements change. Every company modifies its constellation parameters more or less in terms of the number of satellites and orbit altitudes. Also, different companies come up with different constellations. For example, even though SpaceX's, OneWeb's, and Telesat's constellations have inclined orbits combined with polar orbits, they have entirely different orbital characteristics in terms of altitude, inclination, number of orbits, etc. To finally reach distributed beamforming for satellite-to-phone, an appropriate constellation design is indispensable. Take the current Starlink constellation, for example, which gives the satellite a fixed elevation angle. When the satellite is about to travel out of the coverage area of the UE, the UE is already under the coverage of other satellites. Consequently, the UE evaluates each satellite's coverage and chooses the optimal one. But in the case of multiple satellites working together in this paper, the situation will undoubtedly become more complicated. The system needs to select the specific satellite composition according to the received power increment required by the UE, the position of each satellite, and their state information to decide beam switching.

\subsection{System Integration with Other Communications}
Spectrum sharing between different communication systems becomes common as the demands on frequency resources increase. In lower frequency bands, the LEO satellite network needs to share the spectrum with the terrestrial cellular network, while in higher frequency bands, it may need to share the resources with medium orbit or geostationary satellites. Spectrum sharing may cause more interference among different communication systems, so managing the interference between multiple systems and applying resource allocation schemes is necessary. A relevant work was presented in \cite{Integration}, where a general spectrum-sharing framework in satellite and terrestrial networks is introduced, both in the downlink and uplink.

\section{Conclusion and Future Work}
In this work, we first introduced the dependence of the existing LEO satellite network on the ground terminal. Without the terminal, UE, such as standard smartphones, cannot access satellite networks directly. The relevant link budget calculation was then given. A new method named distributed beamforming was proposed to compensate for the insufficient received power. With this technology, LEO satellites can offer the Internet to UE directly, making it possible to access the LEO satellite network from everywhere in the world with only a cell phone. The whole structure of the distributed array was also described in detail to explain how it works. 
In addition, EM simulations verified that extremely distant sources can still interfere with each other. By utilizing constructive interference, distributed beamforming could obtain a higher enhancement of received power than in other ways. Two representative beam coverage cases were presented to show the patterns obtained under different input parameters. 

Finally, to tackle the above challenges listed, we provide some potential further as follows.

\begin{itemize}
\item Synchronize the frequency, phase, and time. Each satellite has its Local Oscillator (LO) working independently. These LOs generated by crystal oscillators typically exhibit variations which may cause a phase drift. A master-slave architecture may work on synchronization. In brief, one satellite first transmits a reference signal to another, afterwards, the receiving satellite use Phase-Locked Loops (PLLs) to lock the signal, thus achieving frequency and phase synchronization. Time alignment could also be attempted by sending each other trigger reference signals. Laser communication within ISLs may provide more help in sharing information and reference signals.

\item Suppress the Doppler effect. The high running speed of LEO satellites yields Doppler shift which may severely affect beamforming performance. Therefore, the influence of the extended frequency offset influences the distributed beamforming performance should be analyzed. Besides, finding solutions, i.e., the Doppler shift compensation techniques, to suppress the effect could be important future work. 

\item Combine with the Reconfigurable Intelligent Surface (RIS). RIS integrates a large number of low-cost passive reflective elements on the plane. It can significantly improve the performance of wireless communication networks by intelligently reconfiguring the wireless propagation environment. Numerous advantages, such as low power consumption, flexible deployment, and almost no extra delay and thermal noise, make it competitive in wireless communication. Thus, the combination of distributed beamforming and RIS would have significant potential. From Section III, it is known that satellite parameters can change the shape of the pattern. Taking advantage of this feature, one of the future works should not only design the coverage pattern but also combine it with RIS to intelligently transmit or reflect the beams according to the needs of user coverage. In the meantime, the requirements of distributed beamforming performance will also guide the constellation design and beams witching management.

\item Apply to the 5G FR2 Band. The frequency we use in this paper belongs to the 5G FR1 Band. Future phones may use the 5G FR2 Band, with a typical frequency of 28 GHz, which can significantly improve bandwidth. But conversely, signals at higher frequencies experience more attenuation. Adopting distributed beamforming technology is a competitive solution because it can provide the closest to the theoretical maximum received power increment. And accordingly, fewer satellites would be needed to cooperate, which saves satellite scheduling resources. Furthermore, working in such a frequency band may interfere with other satellite systems, and finding a smooth way to deal with this problem would be necessary.
\end{itemize}

\ifCLASSOPTIONcaptionsoff
  \newpage
\fi




\begin{thebibliography}{1}

\bibitem{LEO3}
I. del Portillo, B. G. Cameron, E. F. Crawley, ``A technical comparison of three low earth orbit satellite constellation systems to provide global broadband,'' \emph{Acta Astronautica}, vol. 159, pp. 123-135, 2019.
\bibitem{6GVision}
S. Chen, Y. -C. Liang, S. Sun, S. Kang, W. Cheng and M. Peng, ``Vision, Requirements, and Technology Trend of 6G: How to Tackle the Challenges of System Coverage, Capacity, User Data-Rate and Movement Speed,'' \emph{IEEE Wireless Commun.}, vol. 27, no. 2, pp. 218-228, Apr. 2020.
\bibitem{LinkBudget}
O. Popescu, ``Power Budgets for CubeSat Radios to Support Ground Communications and Inter-Satellite Links,'' \emph{IEEE Access}, vol. 5, pp. 12618-12625, 2017.

\bibitem{LYNK}
LYNK GLOBAL, INC., ``Messaging from an Orbital Base Station to Cellular User Equipment Applications with Message Processing Via a Card Operating System,'' U.S. Patent US20210360587A1, Nov. 18, 2021.

\bibitem{CF-mMIMO}
M. Y. Abdelsadek, H. Yanikomeroglu and G. K. Kurt, ``Future Ultra-Dense LEO Satellite Networks: A Cell-Free Massive MIMO Approach,'' \emph{2021 IEEE International Conference on Communications Workshops (ICC Workshops)}, pp. 1-6, 2021.
\bibitem{SatMIMO 16}
B. Ramamurthy, W. G. Cowley and G. Bolding, ``MIMO Applicability to UHF SATCOM,'' \emph{IEEE Global Communications Conference (GLOBECOM)}, pp. 1-7, 2016.
\bibitem{SatMIMO 21}
H. Peng, C. Qi, Y. Chen, Q. Zhang, D. Chen and R. Ding, ``Channel Modeling and Signal Transmission for Land Mobile Satellite MIMO,'' \emph{2021 IEEE Global Communications Conference (GLOBECOM)}, pp. 1-6, 2021.
\bibitem{MISO}
L. Zheng and D. Tse, ``Diversity and multiplexing: a fundamental tradeoff in multiple-antenna channels,'' \emph{IEEE Trans. Inf. Theory}, vol. 49, no. 5, pp. 1073-1096, May 2003.
\bibitem{Beamforming}
P. Ioannides and C. A. Balanis, ``Uniform circular and rectangular arrays for adaptive beamforming applications,'' \emph{IEEE Antennas Wireless Propag. Lett.}, vol. 4, pp. 351-354, 2005.

\bibitem{DTB}
R. Mudumbai, D. R. Brown Iii, U. Madhow and H. V. Poor, ``Distributed transmit beamforming: challenges and recent progress,'' \emph{IEEE Commun. Mag.}, vol. 47, no. 2, pp. 102-110, Feb. 2009.
\bibitem{3GPP}
3GPP, ``5G; NR; User Equipment (UE) radio transmission and reception; Part 1: Range 1 Standalone (3GPP TS 38.101-1 version 17.6.0 Release 17),'' June 2022.
\bibitem{Superposition}
H. G. Schantz, ``On the Superposition and Elastic Recoil of Electromagnetic Waves,'' \emph{FERMAT}, vol. 4, no. 2, 2014.
\bibitem{Optical}
H. Kaushal and G. Kaddoum, ``Optical Communication in Space: Challenges and Mitigation Techniques,'' \emph{Commun. Surveys Tuts.}, vol. 19, no. 1, pp. 57-96, 2017.
\bibitem{Doppler}
I. Ali, N. Al-Dhahir and J. E. Hershey, ``Doppler characterization for LEO satellites,'' \emph{IEEE Trans. Commun.}, vol. 46, no. 3, pp. 309-313, Mar. 1998.
\bibitem{Integration}
C. Zhang, C. Jiang, L. Kuang, J. Jin, Y. He and Z. Han, ``Spatial Spectrum Sharing for Satellite and Terrestrial Communication Networks,'' \emph{IEEE Trans. Aerosp. Electron. Syst.}, vol. 55, no. 3, pp. 1075-1089, June 2019.

\end{thebibliography}
%

%

\begin{IEEEbiographynophoto}{Zhuoao Xu}
received his B.Eng. degree from Harbin Institute of Technology (HIT) in 2015, and his M.Eng. degree in Electronic Engineering from Shanghai Jiao Tong University (SJTU) in 2019. He was rated as the outstanding graduate of Shanghai in 2019. He is currently pursuing his Ph.D. degree with 5G Innovation Centre (5GIC), 6G Innovation Centre (6GIC) and Institute for Communication System (ICS) at the University of Surrey. His current research interests include distributed beamforming, Low Earth Orbit (LEO) constellations, and satellite-to-phone technologies.

\end{IEEEbiographynophoto}

\begin{IEEEbiographynophoto}{Yue Gao}
is a Professor and Chair in Wireless Communications at Institute for Communication Systems, Home of 5GIC and 6GIC, University of Surrey. He develops fundamental research into practice in the interdisciplinary area of smart antennas, signal processing, spectrum sharing, millimetre-wave and internet of things technologies in mobile and satellite systems. He has published over 200 peer-reviewed journal and conference papers, one book and five book chapters. He is an Engineering and Physical Sciences Research Council Fellow from 2018 to 2023.
\end{IEEEbiographynophoto}

\begin{IEEEbiographynophoto}{Gaojie Chen}
Gaojie Chen (Senior Member, IEEE) received the B.Eng. and B.Ec. degrees in electrical information engineering and international economics and trade from Northwest University, Xi’an, China, in 2006, and the M.Sc. (Hons.) and Ph.D. degrees in electrical and electronic engineering from Loughborough University, Loughborough, U.K., in 2008 and 2012, respectively. He is currently an Assistant Professor with the Institute for Communication Systems, 5GIC \& 6GIC, University of Surrey. His research interests include information theory, wireless communications, cooperative communications, cognitive radio, Internet of Things, secrecy communications, and random geometric networks.
\end{IEEEbiographynophoto}

\begin{IEEEbiographynophoto}{Ryan Fernandez}
received his BSc degree in Physics at the University of Surrey in 2018 and his MSc degree in Electronic Engineering with specialisation in Space Engineering at the University of Surrey in 2020. He is currently reading towards his PhD with the 5G Innovation Centre (5GIC), 6G Innovation Centre (6GIC) and Institute for Communication Systems (ICS) at the University of Surrey. His current research interests include Low Earth Orbit (LEO) mega constellations and direct satellite to mobile technologies.
\end{IEEEbiographynophoto}


\begin{IEEEbiographynophoto}{Vedaprabhu Basavarajappa}
earned his PhD from the University of Cantabria, Spain in 2018. He obtained his MS from TU Delft, the Netherlands in 2012 in Electrical Engineering with specialization in Telecommunication and his Bachelor’s degree in Electronics and Communication from Visvesvaraya Technological University, India. Following this he was a Postdoctoral Researcher at the 5G Innovation Centre (5GIC/ 6GIC/ ICS) at the University of Surrey, United Kingdom. His research interests include 5G antennas, mmwave, satellite antennas and phased array antennas.
\end{IEEEbiographynophoto}

\begin{IEEEbiographynophoto}{Rahim Tafazolli}
is Regius Professor of Electronic Engineering, Professor of Mobile and Satellite Communications, Founder and Director of 5G Innovation Centre (5GIC), 6G Innovation Centre (6GIC) and Institute for Communication System (ICS) at the University of Surrey. He has over 30 years of experience in digital communications research and teaching. He has authored and co-authored more than 1000 research publications and is regularly invited to deliver keynote talks and distinguished lectures to international conferences and workshops. He has been named a Fellow of the Royal Academy of Engineering in 2020.
\end{IEEEbiographynophoto}




\end{document}